\documentclass[journal]{IEEEtran}

\ifCLASSINFOpdf
\else
   \usepackage[dvips]{graphicx}
\fi
\usepackage{url}
\usepackage{graphicx}
\usepackage{hyperref}
\usepackage{lineno}
\usepackage{stfloats}
\usepackage{enumitem}
\usepackage{amsmath}
\usepackage{cases}
\usepackage{tipa}
\usepackage[T1]{fontenc}

\newtheorem{theorem}{Theorem}
\newtheorem{definition}{Definition}
\newtheorem{lemma}{Lemma}

\begin{document}

\title{Adjacency-hopping de Bruijn Sequences for Non-repetitive Coding}

\author{Bin~Chen,~Zhenglin~Liang,~Shiqian~Wu,~\IEEEmembership{Senior Member,~IEEE}
\thanks{(Corresponding author: Shiqian Wu.)
	
	Bin Chen, Zhenglin Liang, and Shiqian Wu are with the Institute of Robotics and Intelligent Systems, School of Information Science and Engineering, Wuhan University of Science and Technology, Wuhan, 430081, China, and Shiqian Wu is also with the Hubei Province Key Laboratory of Intelligent Information Processing and Real-Time Industrial System, Wuhan, 430072, China (e-mail: chenbin@wust.edu.cn; yeasukura3@gmail.com; shiqian.wu@wust.edu.cn).}

}

\markboth{Journal of \LaTeX\ Class Files, Vol. 14, No. 8, August 2015}
{Shell \MakeLowercase{\textit{et al.}}: Bare Demo of IEEEtran.cls for IEEE Journals}
\maketitle

\begin{abstract}
A special type of cyclic sequences named adjacency-hopping de Bruijn sequences is introduced in this paper. It is theoretically proved the existence of such sequences, and the number of such sequences is derived. These sequences guarantee that all neighboring codes are different while retaining the uniqueness of subsequences, which is a significant characteristic of original de Bruijn sequences in coding and matching. At last, the adjacency-hopping de Bruijn sequences are applied to structured light coding, and a color fringe pattern coded by such a sequence is presented. In summary, the proposed sequences demonstrate significant advantages in structured light coding by virtue of the uniqueness of subsequences and the adjacency-hopping characteristic, and show potential for extension to other fields with similar requirements of non-repetitive coding and efficient matching.
\end{abstract}


\IEEEpeerreviewmaketitle

\section{Introduction}\label{sec_one}

\IEEEPARstart{D}{e} Bruijn sequences are first named after N. G. de Bruijn who reported de Bruijn sequences with only two (binary) codes (symbols) in \cite{deBruijn1946}. In \cite{deBruijn1951}, $k$-code de Bruijn sequences are generalized from the original binary de Bruijn sequences. The $n$-order de Bruijn sequence with $k$ codes is a cyclic sequence in which each element belongs to a code set of size $k$ and every subsequence of length $n$ occurs only once. Since the uniqueness of the subsequences, length-$n$ subsequences matching on a de Bruijn sequence is efficient, i.e. the computational complexity of the naive sequential matching is $O(n)$ and some advanced matching algorithms (e.g. Hash-based matching \cite{chikhi2013space} and Look-up table matching \cite{zhang2021timing}) achieve $O(1)$ ideally. Due to this significant advantage, de Bruijn sequences are applied for coding and matching in various fields (e.g. structured light coding \cite{zhang2002,pages2004new,petkovic2016}, genome sequencing \cite{compeau2011apply}, etc.). However, a limitation arises from the presence of identical consecutive codes within a de Bruijn sequence, which can yield challenges in certain specialized applications that require non-repetitive coding. An example of such a scenario is when a structured light pattern is encoded using a de Bruijn sequence, it becomes difficult to accurately determine the number and precise positions of fringes within the same color areas in captured images. Obviously, inserting a narrow gap by an additional color into all neighboring fringes is a straightforward solution \cite{pages2004new,petkovic2016}. However, this straightforward approach is not robust due to occlusion situations in practice. Zhang et al. propose another solution in \cite{zhang2002}, in which a de Bruijn sequence is transformed into another sequence without the repetitive neighboring codes by a recursive binary $XOR$ operation. Nevertheless, the transformed sequence is no longer a de Bruijn sequence which yields low-efficient matching.

In this paper, a type of restricted de Bruijn sequences named adjacency-hopping de Bruijn sequences is proposed. Such sequences guarantee that all neighboring codes are different while holding the uniqueness of the subsequences. The rest part of this paper is as follows. After defining the adjacency-hopping de Bruijn sequences, the generation is given first and the existence of these sequences is then proved. At last, the number of $k$-code $n$-order adjacency-hopping sequences is computed following several necessary lemmas.

\section{Adjacency-hopping de Bruijn Sequences}\label{sec_two}
\subsection{Definition}\label{sec_defBS}
There are several definitions of de Bruijn sequences in different subjects (e.g. graph theory, discrete mathematics and combinatorics, etc.). The following equivalent definition is introduced in this study.

\begin{definition}[de Bruijn Sequences]\label{def_BS}
	{A de Bruijn sequence of order $n$ on a size-$k$ code set $\mathbf{M}$, which is denoted as $B(k,n)$, is a cyclic sequence in which every possible length-$n$ sequence on $\mathbf{M}$ occurs exactly once as a subsequence.}
\end{definition}

Then the precise definition of the adjacency-hopping de Bruijn sequences is given below.

\begin{definition}[Adjacency-hopping de Bruijn Sequences]\label{def_AHBS}
	{A sequence in which every code is different to its neighboring codes is an adjacency-hopping sequence. Then an adjacency-hopping de Bruijn sequence of order $n$ on a size-$k$ code set $\mathbf{M}$, which is denoted as $H(k, n)$, is a cyclic and adjacency-hopping sequence in which every possible length-$n$ adjacency-hopping subsequence on $\mathbf{M}$ occurs only once.}
\end{definition}

It is obvious to find the differences between de Bruijn sequences and adjacency-hopping ones: 

\begin{enumerate}
	\item All neighboring codes in an adjacency-hopping de Bruijn sequence are different, but the original de Bruijn sequence does not hold.
	\item For a same code set $\mathbf{M}$, the number of possible subsequences in $H(k,n)$ is smaller than $B(k,n)$'s. Thus the length of $H(k,n)$ is shorter as well.
\end{enumerate}

\subsection{Generation and Existence}\label{sec_existence}
As discussed in \cite{deBruijn1946,deBruijn1951,Alge2013}, there is a correspondence between de Bruijn sequences and Eulerian tours in a certain directed graph, which is referred to the de Bruijn graph. In other words, if an Eulerian tour on the $k$-code $n$-order de Bruijn graph is found out, a $k$-code $n$-order de Bruijn sequence can be generated from it. As a restricted de Bruijn sequence, an adjacency-hopping de Bruijn sequence can be generated in a similar way as follows.
\begin{definition}[Corresponding Graph]\label{def_CORRG}
	{The corresponding graph $G_k^n=(\mathbf{V}, \mathbf{E})$ of a $k$-code $n$-order adjacency-hopping de Bruijn sequence $H(k,n)$ ($k\geq 2, n\geq 2$) on a size-$k$ code set $\mathbf{M}$ is a directed graph where:}
	\begin{itemize}
		\item $\mathbf{V}=\{v_1v_2\cdots v_i \cdots v_{n-1}|\forall(1 \leq i < n-1), (v_i, v_{i+1} \in \mathbf{M}) \wedge (v_i\neq v_{i+1})\}$ { is a set of all possible vertices}. 
		\item $\mathbf{E}=\{(u,v)|u, v\in \mathbf{V}\}$ { is an edge set where there is an edge} $e=(u, v)$ { from vertex} $u=u_1u_2\cdots u_{n-1}$ { to vertex} $v=v_1v_2\cdots v_{n-1}$ { if and only if} $u_j=v_{j-1}$ { for each} $2\leq j \leq n-1$.
	\end{itemize}
\end{definition}

\begin{figure}
	\centerline{\includegraphics[width=0.65\linewidth]{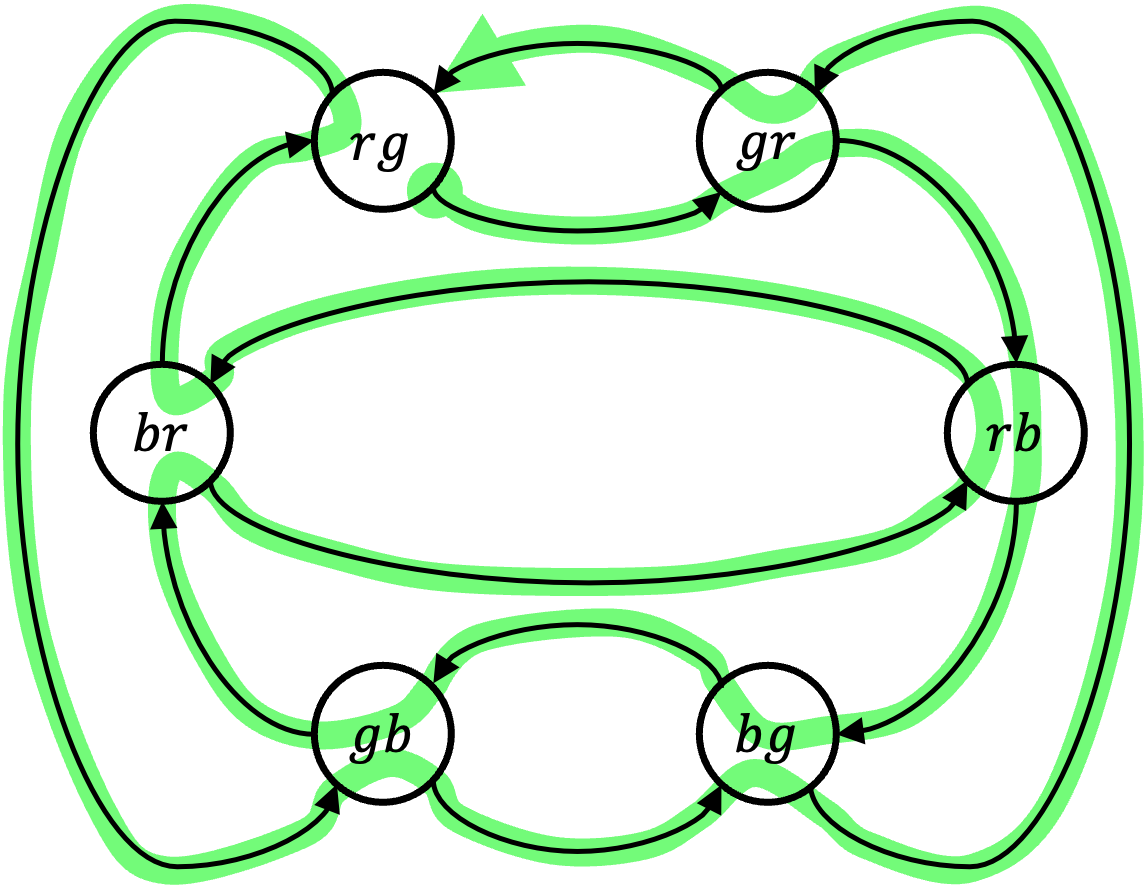}}
	\caption{The corresponding graph $G_3^3$ on $\mathbf{M}=\{r, g, b\}$. The path demonstrated by a green curve is an Eulerian tour.}\label{fig_graph}
\end{figure}

Compared with the definition of de Bruijn graphs in \cite{deBruijn1946,deBruijn1951,Alge2013}, the corresponding graph $G_k^n$ is a subgraph of the $k$-code $n$-order de Bruijn graph. Let $\mathcal{T}(e) = v_{n-1} $ in which $e$ denotes an edge as Definition \ref{def_CORRG}. If an Eulerian tour $\mathcal{E}=e_1 e_2\cdots $ exists on $G_k^n$, an adjacency-hopping sequence associated with the Eulerian tour $\mathcal{E}$ can be generated as $\mathcal{T}(e_1) \mathcal{T}(e_2) \cdots$.  For instance, Fig. \ref{fig_graph} shows the corresponding graph $G_3^3$ on a code set $\mathbf{M} = \{r, g, b\}$. According to the Definition \ref{def_CORRG}, the graph $G_3^3$ has six vertices($rg$, $gr$, $rb$, $bg$, $gb$ and $br$) and twelve edges. An Eulerian tour demonstrated by a green curve in Fig. \ref{fig_graph} is as follows (only the vertices passed through are listed in sequence):

\begin{equation*}
	\underset{\cdots}{\underline{\phantom{\underline{\underline{rb, bg}}}}}
	\llap{$\underset{e_2}{\underline{\phantom{\underline{gr, rb}}}}, \phantom{bg}$}
	\llap{$\underset{e_1}{\underline{rg, gr}}, rb, bg$}
	, gb, br, rb, br, rg, gb, bg, gr, rg
\end{equation*}

Thus an adjacency-hopping sequence associated with this Eulerian tour is $rbgbrbrgbgrg$ (i.e. $\mathcal{T}(e_1)=r$, $\mathcal{T}(e_2)=b$, \\etc.). It is easy to find that every possible adjacency-hopping subsequence of length 3 in $\mathbf{V}=\{rbg, bgb, gbr, brb, rbr, brg,$\\$ rgb, gbg, bgr, grg, rgr, grb\}$ occurs exactly only once, and all neighboring codes in the generated sequence are different. We will give an elegant proof of the existence of adjacency-hopping sequences in the following.

\begin{lemma}\label{lemma_adj}
	{Let $\mathbf{A}$ be the directed adjacency matrix of $G_k^n = (\mathbf{V},\mathbf{E})$. If the rows and columns are indexed by the vertices, the $(u,v)$-entry of $\mathbf{A}$ is given by}
	\begin{equation*}
		\mathbf{A}_{uv} = \begin{cases}
			1, & \text{{ if}} (u,v)\text{{  is an edge;}} \\
			0, & \text{{ Otherwise.}}
		\end{cases}
	\end{equation*}
	{Then $\mathbf{A}^{n-1}+\mathbf{A}^{n-2}=\mathbf{J}$, where $\mathbf{J}$ is an all $1$'s matrix with the same size of $\mathbf{A}$.}
\end{lemma}

\begin{IEEEproof}
	{The constraint of the edge set $\mathbf{E}$ in Definition $\ref{def_CORRG}$ is equivalent to the statement that, for every edge in $\mathbf{E}$, the length-$(n-2)$ prefix of the start vertex is equal to the length-$(n-2)$ suffix of the end vertex. Therefore, suppose two vertices $u=u_1u_2\cdots u_{n-1}$ and $v=v_1v_2\cdots v_{n-1}$ in the vertex set $\mathbf{V}$, we have:}
	\begin{enumerate}
		\item {If $u_{n-1} \neq v_1$, i.e. the last code of $u$ does not equal to the first code of $v$, the  unique path of length $n-1$ from $u$ to $v$ is} 
		\begin{equation*}
			\begin{aligned}
				&\underset{e_1}{\underline{(u_1u_2\cdots u_{n-1}, u_2\cdots u_{n-1}v_1)} }\to \\
				&\underset{e_2}{\underline{(u_2\cdots u_{n-1}v_1, u_3\cdots u_{n-1}v_1v_2)}} \to \cdots \to \\
				&\underset{e_{n-1}}{\underline{(u_{n-1}v1\cdots v_{n-2},v_1v_2\cdots v_{n-1})}}
			\end{aligned}			
		\end{equation*}         
		{and there are no paths of length $n-2$ from $u$ to $v$. Thus, $\mathbf{A}^{n-1}_{uv} =1$ and $\mathbf{A}^{n-2}_{uv} =0$.}
		\item {Otherwise, if $u_{n-1} = v_1 = m$, i.e. the last code of $u=u_1u_2\cdots m$ equals to the first code of $v = m v_2\cdots v_{n-1}$, there are no paths of length of $n-1$ and the unique path of length $n-2$ from $u$ to $v$ is} 
		\begin{equation*}
			\begin{aligned}
				&\underset{e_1}{\underline{(u_1u_2\cdots m, u_2\cdots m v_2)} }\to \\
				&\underset{e_2}{\underline{(u_2\cdots m v_2, u_3\cdots m v_2v_3)}} \to \cdots \to \\
				&\underset{e_{n-2}}{\underline{(u_{n-2} m v_2\cdots v_{n-2},m v_2\cdots v_{n-1})}} 
			\end{aligned}
		\end{equation*}
		{then we have $\mathbf{A}^{n-2}_{uv} = 1$ and $\mathbf{A}^{n-1}_{uv} = 0$.}
	\end{enumerate}
	
	{Combining two cases above, then each element of the result of $\mathbf{A}^{n-1}+\mathbf{A}^{n-2}$ is $1$, and the proof follows.}
\end{IEEEproof}

\begin{theorem}\label{the_Existance}
	For a size-$k$ code set $\mathbf{M}$, except for $k=1$ and $n>1$, there exists at least a cyclic sequence $H(k,n)$ satisfying Definition $\ref{def_AHBS}$ for all positive integers $n$ and $k$, and the length of the sequence $H(k,n)$ is $k(k-1)^{n-1}$.
\end{theorem}

\begin{IEEEproof}
	
	{Case 1: $k=1$ and $n>1$. Assuming that a sequence $H(1,n)$ exists, it follows immediately that every code in $H(1, n)$ is the same, which contradicts the definition of adjacency-hopping de Bruijn sequences (See Definition $\ref{def_AHBS}$). Therefore, $H(1,n)$ does not exist when $k=1$ and $n>1$.}\label{item:case1}
	
	{Case 2: $k\geq 1$ and $n=1$. A first-order adjacency-hopping de Bruijn sequence $H(k,1)$ can be formed by permuting the $k$ codes. Thus, the length of $H(k,1)$ equals to $k$.}\label{item:case2}
	
	{Case 3: $k\geq 2$ and $n\geq 2$. The corresponding graph $G_k^n$ associated with $H(k,n)$ exists. For any two fixed vertices $u$ and $v$ in the corresponding graph $G_k^n$ ($n \geq 2, k \geq 2$), the Lemma $\ref{lemma_adj}$ is equivalent to the assertion that there is at least a path with length $n-2$ or $n-1$ from $u$ to $v$. Thus $G_k^n$ is connected.}\label{item:case3}
	
	{Now let $\delta^-_u$ and $\delta^+_u$ denote the indegree and outdegree of the vertex $u$, respectively. By the Definition $\ref{def_CORRG}$, it is easy to find that there are $k-1$ directed edges from $u$ to other $k-1$ different vertices. Similarly there are $k-1$ directed edges from $k-1$ different vertices to $u$. Hence $\delta^+_u = \delta^-_u = k-1$, namely, the graph $G_k^n$ is directed and balanced equivalently. Accordingly, $G_k^n$ has at least an Eulerian tour due to its balance and connectivity. Equivalently, an adjacency-hopping de Bruijn sequence $H(k,n)$ exists.}
	
	{Furthermore, the number of edges (or the length of Eulerian tours) in $G_k^n$ is equal to $k(k-1)^{n-1}$, since $G_k^n$ has $k(k-1)^{n-2}$ vertices and the outdegree of each vertex is $k-1$. Thus the length of the sequence $H(k,n)$ is $k(k-1)^{n-1}$, which is satisfied in the Case 2 as well, and the proof follows.}
\end{IEEEproof}

From the proof procedure above, some properties of the corresponding graph $G_k^n$ can be summarized as follows:
\begin{enumerate}
	\item The outdegree and indegree of all vertices in $G_k^n$ are equal to $k - 1$.   
	\item There are $k(k-1)^{n-2}$ vertices and $k(k-1)^{n-1}$ edges in $G_k^n$.
	\item The graph $G_k^n$ is a directed graph which is connected, balanced and loopless (no edge from a vertex to itself in $G_k^n$). 
\end{enumerate}

\subsection{Enumeration}\label{sec_enum}
In this section, we compute enumeration of adjacency-hopping sequences by counting the number of Eulerian tours on the corresponding graph. First, eigenvalues of the directed adjacency matrix of the $k$-code $n$-order corresponding graph are computed as below.

\begin{lemma}\label{lemma_EIG}
	{Assuming $\mathbf{A}$ is the directed adjacency matrix of $G_k^n$, eigenvalues of $\mathbf{A}$ are $k-1$ (with multiplicity 1), $-1$ (with multiplicity $k-1$) and $0$ (with multiplicity $k(k-1)^{n-2}-k$).}  
\end{lemma}

\begin{IEEEproof}
	{Let $\ell$ be the number of vertices of $G_k^n$. The directed adjacency matrix $\mathbf{A}$ is a $\ell \times \ell$ matrix. Suppose that $\lambda_1, \cdots \lambda_i, \cdots, \lambda_\ell$ are eigenvalues of $\mathbf{A}$, so $\lambda_1^c, \cdots \lambda_i^c, \cdots, \lambda_\ell^c$ are eigenvalues of $\mathbf{A}^c$ for all positive integers $c$. Hence, we have $(\mathbf{A}^{n-1} + \mathbf{A}^{n-2})x = (\lambda_i^{n-1} + \lambda_i^{n-2}) x$, where $x$ is the eigenvector associated with $\lambda_i$. Thus $\lambda_i^{n-1} + \lambda_i^{n-2}$ is the eigenvalue of the summation of $\mathbf{A}^{n-1}$ and $\mathbf{A}^{n-2}$. Moreover, by Lemma $\ref{lemma_adj}$, we have $\mathbf{A}^{n-1} + \mathbf{A}^{n-2} = \mathbf{J}$ where $\mathbf{J}$ is a $\ell \times \ell$ matrix of all $1$'s. It can be deduced that $\mathbf{J}$ has an eigenvalue $\ell$ and $\ell - 1$ eigenvalues equal to $0$.} 
	
	{Now assuming that the last eigenvalue of $\mathbf{J}$ is $\ell$, we have}
	\begin{subequations}
	\begin{numcases}{}
		\lambda_i^{n-2}\left(\lambda_i+1\right)=0, i=1,2, \cdots \ell-1;\label{eq:numcase1}\\
		\lambda_{\ell}^{n-2}\left(\lambda_{\ell}+1\right)=\ell.\label{eq:numcase2}
	\end{numcases}
	\label{eq}
\end{subequations}
	{Put $\ell = k(k-1)^{n-2}$ (See the second property of $G_k^n$ in $\ref{sec_existence}$) in the right side of \eqref{eq:numcase2}, then the eigenvalues of $\mathbf{A}$ can be deduced as follows:}
	\begin{equation}\label{equ_lemma_EIG_2}
	\begin{cases}
		\lambda_i = 0 \text{ or } -1, \text{  }i = 1, 2, \cdots \ell -1;\\
		\lambda_\ell = k-1.
	\end{cases}
\end{equation}
{Since $G_k^n$ is loopless, the trace of $\mathbf{A}$ equals to $0$, i.e. $tr(\mathbf{A}) =\sum_{i=1}^{k}\lambda_i = 0$. Combining \eqref{equ_lemma_EIG_2}, we have $\lambda_\ell = k-1$, $k-1$ of $\lambda_1, \cdots \lambda_{\ell - 1}$ are equals to $-1$ and the rest $\ell - k$ eigenvalues are $0$. Hence, setting $\ell=k(k-1)^{n-2}$ yields the desired assertion.}
\end{IEEEproof}

To set up the correspondence between the number of Eulerian tours on a connected directed graph and the eigenvalues of its adjacency matrix, the following lemma, which is deduced from the BEST Theorem in \cite{deBruijn1951} and the Matrix-Tree Theorem in \cite{Alge2013}, is introduced first.

\begin{lemma}[Corollary 10.5 in \cite{Alge2013}]\label{lemma_alge2013}
	{Let $G$ be a connected balanced directed graph with $\ell$ vertices, $u$, $v$ be two vertices in $G$ and $\delta_u^+$ be outdegree of $u$, and $\mathbf{L}$ be a $\ell\times \ell$ matrix defined by}
	\begin{equation*}
		\mathbf{L}_{uv}=\begin{cases}
			-q_{uv}, & \text { {if }} u\neq v \text{ { and there are }}q_{uv} \\ 
			&\text{ { edges from vertex }}u\text{ { to vertex }}v;\\
			\delta_u^+ - I_u, & \text { {if }} u = v \text{ { and there are }}I_u \\ 
			&\text{ { edges from vertex }}u \text{ { to itself}} { (loop).}
		\end{cases}
	\end{equation*}
	{Let $e$ be an edge of $G$, and let $\#(G,e)$ denote the number of Eulerian tours of $G$ with first edge $e$. If $\mathbf{L}$ has eigenvalues $\lambda_1, \cdots, \lambda_\ell$ with $\lambda_\ell = 0$, then }
	\begin{equation}\label{eqn_er2013_num}
		\#(G,e) = \frac{1}{\ell}\lambda_1\cdots \lambda_{\ell-1}\prod_{u\in \mathbf{V}}(\delta^+_u -1).
	\end{equation}
\end{lemma}

By Lemma \ref{lemma_EIG} and Lemma \ref{lemma_alge2013}, the number of Eulerian tours on $k$-code $n$-order corresponding graph is given as follows.
\begin{theorem}\label{ER_num}
	{Let $G_k^n$ be the corresponding graph of a $k$-code $n$-order adjacency-hopping de Bruijn sequence, and $\#(G_k^n,e)$ denote the number of Eulerian tours of $G_k^n$ with first edge $e$. Then} 
	\begin{equation}\label{eqn_er_num}
		\#(G_k^n,e) = \left(\frac{k}{k-1}\right)^{k-2}\frac{[(k-1)!]^{k(k-1)^{n-2}}}{(k-1)^n}.
	\end{equation}
\end{theorem}

\begin{IEEEproof}
	{Let $\mathbf{L}$ denote as Lemma $\ref{lemma_alge2013}$. According to the Definition $\ref{def_AHBS}$ and properties of $G_k^n$, we have} 
	\begin{itemize}
		\item { $G_k^n$ is a connected balanced directed graph (loopless),}
		\item { there is at most only one edge from a fixed vertex to another fixed one,}
		\item { the outdegree $\delta_u^+$ of a vertex $u$ is $k-1$,}
		\item { the number $\ell$ of vertices in $G_k^n$ equals to $k(k-1)^{n-2}$.}
	\end{itemize}
	{ Therefore $\mathbf{L}$ follows}
	\begin{equation*}
		\mathbf{L}_{uv}=
		\begin{cases}
			k-1, & \text{{ if }} u = v; \\
			-1, & \text{{ if }} u \neq v \text{ {  and there is an edge}}\\ 
			&\text{{ from vertex $u$ to $v$}};\\
			0, & \text{{ otherwise }}.
		\end{cases}
	\end{equation*} 
	{Then we have $\mathbf{L} = (k - 1)\mathbf{I} - \mathbf{A}$ in which $\mathbf{I}$ denotes the identify matrix and $\mathbf{A}$ is the directed adjacency matrix of $G_k^n$.}
	
	{Now assume that $\lambda$ is an eigenvalue of $\mathbf{A}$, so $(k-1) - \lambda$ is an eigenvalue of $\mathbf{L}$. By Lemma $\ref{lemma_EIG}$, $\mathbf{A}$ has eigenvalues $k-1$ with multiplicity $1$, $-1$ with multiplicity $k-1$ and $0$ with multiplicity $\ell-k$. Thus, the eigenvalues of $\mathbf{L}$ are 0 (with multiplicity $1$), $k$ (with multiplicity $k-1$) and $k-1$ (with multiplicity $\ell-k$).} 
	
	{Finally, let $\#(G_k^n, e)$ denote the number of Eulerian tours of $G_k^n$ with first edge $e$. Putting $\delta_u^+ = k-1$ and all eigenvalues of $\mathbf{L}$ in \eqref{eqn_er2013_num}, we have} 
	\begin{eqnarray*}
		\#(G_k^n, e) &=& \frac{1}{\ell}\prod_{1}^{k-1}k \prod_{1}^{\ell-k}{(k-1)} \prod_{1}^{\ell}(k-2)!\\
		&=& \frac{1}{\ell}k^{k-1}(k-1)^{\ell-k}[(k-2)!]^\ell\\
		&=& \frac{k^{k-1}}{\ell(k-1)^{k}}[(k-1)!]^\ell.  \\      
	\end{eqnarray*}
	{Putting $\ell = k(k-1)^{n-2}$ in the formula above, $\#(G_k^n, e) = \left(\frac{k}{k-1}\right)^{k-2}\frac{[(k-1)!]^{k(k-1)^{n-2}}}{(k-1)^n}$ as desired.}
\end{IEEEproof}

By the definitions, lemmas and theorems above, the number of $k$-code $n$-order adjacency-hopping sequences is given in the following:
\begin{theorem}\label{AHBS_num}
	Let $\#H_k^n$ denote the number of $k$-code $n$-order adjacency-hopping de Bruijn sequences $H(k,n)$. Then $\#H_k^n$ follows
	\begin{equation}\label{eqn_ahbs_num}
		\#H_k^n=\begin{cases}
			(k-1)!, & k\geq1 \text{ and } n=1,\\
			\left(\frac{k}{k-1}\right)^{k-2}\frac{[(k-1)!]^{k(k-1)^{n-2}}}{(k-1)^n}, & k\geq2 \text{ and } n\geq 2.
		\end{cases}
	\end{equation}	
\end{theorem}

\begin{IEEEproof}
	Since adjacency-hopping sequences are cyclic, it is easily found that if $m_1 m_2 \cdots m_i \cdots m_q$ is an adjacency-\\hopping sequence, all sequences in the set $\{m_i \cdots m_q m_1 \cdots $\\$m_{i - 1}| 1 < i \leq q\}$ are also adjacency-hopping sequences, and these sequences are equivalent. Therefore, such sequences are counted only once for the enumeration of adjacency-hopping sequences. In other words, the number of adjacency-hopping sequences is actually independent of the first code. Hence, $\#H_k^n$ can be computed out by enumerating permutations of $k$ codes with a fixed initial code in the case $k\geq 1$, $n=1$, or by enumerating Eulerian tours of the corresponding graph $G_k^n$ with a fixed first edge in the case $k\geq 2$, $n\geq 2$. By combining the proof of Theorem $\ref{the_Existance}$ and the Theorem $\ref{ER_num}$, we obtain \eqref{eqn_ahbs_num}, as claimed.
\end{IEEEproof}

\subsection{Example: Fringe Pattern Coding}
\begin{figure}
	\centerline{\includegraphics[width=1.0\linewidth]{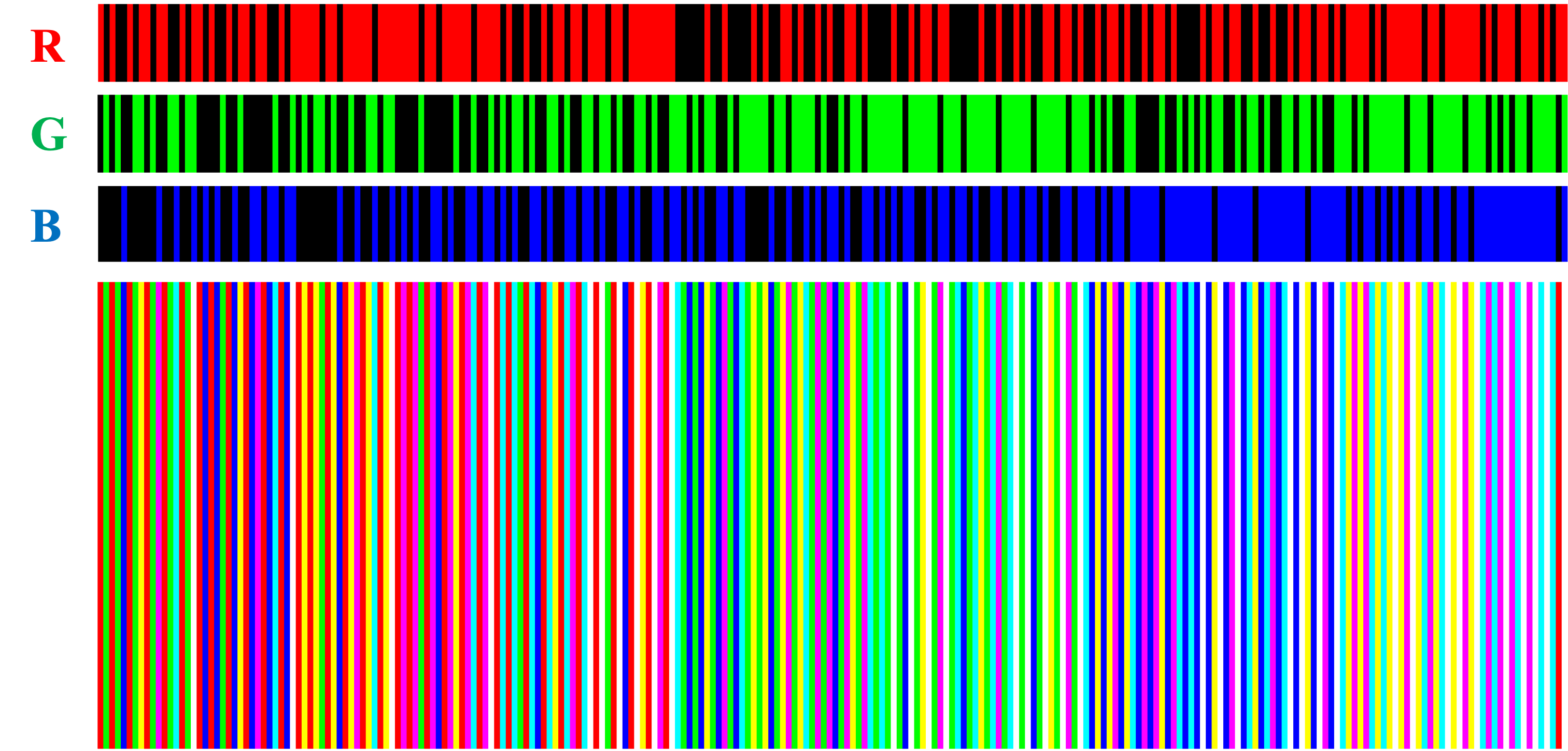}}
	\caption{A color fringe pattern base on a $H(7, 3)$ sequence with $\mathbf{M}=\{(0, 0, 1), (0, 1, 0), (0, 1, 1), (1, 0, 0), (1, 0, 1), (1, 1, 0), (1, 1, 1)\}$.}\label{fig_pattern}
\end{figure}

Here we give a color fringe pattern coded by adjacency-hopping sequences as an example. Seven of eight binary combinations of $(R, G, B)$ colors, in which black $(0, 0, 0)$ is eliminated, are applied to generate a color fringe pattern. Fig. \ref{fig_pattern} presents a color fringe pattern coded by a $7$-code $3$-order adjacency-hopping de Bruijn sequence $H(7, 3)$ on a code set  $\mathbf{M}=\{(0, 0, 1), (0, 1, 0), (0, 1, 1), (1, 0, 0), (1, 0, 1), (1, 1, 0),$\\$ (1, 1, 1)\}$ in which each code corresponds to a color combined by red, green and blue, i.e. blue, green, yellow and so on. Moreover, the sequence applied in Fig. \ref{fig_pattern} is generated by the Hierholzer's algorithm \cite{ozcan2019continuous} which is a well-known Eulerian tours searching algorithm in graph theory. 

In practice, color fringe patterns with different number of fringes can be generated by selecting appropriate $k$ and $n$ or by truncating from a longer sequence. For example, an adjacency-hopping de Bruijn sequence $H(7, 3)$ applied in Fig. \ref{fig_pattern}, has 252 codes according to Theorem \ref{the_Existance}, and it can be truncated into a sequence shorter than $252$ which still hold the uniqueness of the subsequences and the adjacency-hopping characteristic.

\section{Conclusion}\label{sec_three}
In this paper, we defined a type of restricted de Bruijn sequences derived from the original de Bruijn sequences. In such a sequence, every possible subsequence with a certain length occurs only once and all neighboring codes are different. Therefore, we named it as adjacency-hopping de Bruijn sequences. By defining a corresponding graph, we set up a correspondence between $k$-code $n$-order adjacency-hopping sequences and a connected balanced directed graph. Then the fundamental problems, i.e. existence, generation and enumeration, of such sequences are transformed to the existence, generation and enumeration problems of Eulerian tours on the corresponding graph. Based on the works above and combined with necessary principles in other related subjects, not only the existence of adjacency-hopping sequences is proved, but also the formula for the number of such sequences is deduced from several theorems and lemmas. At last, the structured light coding is selected as a typical non-repetitive coding example and a color fringe pattern coded by an adjacency-hopping sequence is illuminated. In summary, due to the uniqueness of the subsequences and the adjacency-hopping characteristic, adjacency-hopping de Bruijn sequences have significant advantages in structured light coding and could be extended to other similar topics and beyond.

\end{document}